\documentclass[aps,prd,twocolumn,
superscriptaddress,floatfix,showpacs,showkeys,preprintnumbers]{revtex4}
\usepackage{graphicx}
\usepackage{amsmath}
\usepackage{amssymb}
\usepackage[hypertex,colorlinks=true,linkcolor=red,citecolor=blue]{hyperref}
\newcounter{comment}

{\refstepcounter{comment}%
\begin{quote}
\ttfamily\small$\blacksquare$ \textbf{\underline{Comment} $\sharp$\thecomment:}}%
{\end{quote}}

{
\begin{quote}
\ttfamily\small$\blacktriangleright$ \textbf{\underline{Reply} $\sharp$\thecomment:}}%
{\end{quote}}

\renewcommand{\theequation}{\arabic{section}.\arabic{equation}}%

\newcommand{\cQ}{{\cal Q}}
\newcommand{\xB}{x_{\rm B}}
\newcommand{\HDD}{H_{\rm DD}}

\newcommand{\be}{\begin{eqnarray}}
\newcommand{\ee}{\end{eqnarray}}
\newcommand{\ba}{\begin{array}}
\newcommand{\ea}{\end{array}}

\newcommand{\bi}{\begin{itemize}}
\newcommand{\ei}{\end{itemize}}

\begin{document}

\title{$J=0$ fixed pole  and  $D$-term form factor in  deeply virtual Compton scattering}

\author{D.~Mueller}
\affiliation{Department of Physics, University of Cape Town, Private Bag X3\\
     7701 Rondebosch, South Africa}

\author{K.~M.~Semenov-Tian-Shansky}
\affiliation{IFPA, d{\'e}partement AGO, Universit{\' e} de Li{\'e}ge,
BE-4000 Li{\' e}ge, Belgium}
\affiliation{CPhT, \'{E}cole Polytechnique, CNRS, FR-91128 Palaiseau, France}

\begin{abstract}
\noindent
S.~Brodsky, F.~J.~Llanes-Estrada, and A.~Szczepaniak emphasized the importance of the
$J=0$
fixed pole manifestation in real and (deeply) virtual Compton scattering measurements
and argued that the $J=0$ fixed pole is universal, {\it i.e.}, independent on the photon virtualities \cite{Brodsky:2008qu}.
In this paper we review the $J=0$ fixed pole issue in  deeply virtual Compton scattering.
We employ the dispersive approach to derive the sum rule that connects the
$J=0$
fixed pole contribution and the subtraction constant, called the $D$-term form factor for deeply
virtual Compton scattering. We show that in the Bjorken limit the
$J=0$
fixed pole universality hypothesis is equivalent to the conjecture that the
$D$-term form factor is given by the inverse moment sum rule for the Compton form factor.
This implies that the $D$-term is an inherent part of corresponding
generalized parton distribution (GPD).
Any supplementary $D$-term added to a GPD results in an additional
$J=0$ fixed pole contribution and implies the violation of the universality hypothesis.
We argue that there exists no theoretical proof for the $J=0$ fixed pole universality conjecture.
\end{abstract}



\maketitle

\section{Introduction}

Compton scattering off a nucleon
\begin{equation}
\gamma^{(\ast)}(q_1) +  N(p_1) \to \gamma^{(\ast)}(q_2) + N(p_2)
\label{ComptonScat}
\end{equation}
with real  photons
($q_{1}^2=q_{2}^2=0$),
with one virtual space-like and one real photon
($q_{1}^2= - Q_1^2 <0$, $q_{2}^2=0$),
and with two virtual space-like photons
($q_{1}^2=-Q_1^2 <0$, $q_{2}^2=-Q_2^2 <0$ )
are important processes to address the internal structure of nucleons from the
low energy to the high energy regime. Depending on the resolution scale which is
set up experimentally, different theoretical frameworks are appropriate
to analyze experimental data and to provide interpretation in terms of the proper degrees of freedom.

Experimental and theoretical investigations of real Compton scattering are going back to
pre-QCD time  and were mainly based on the dispersive approach and the SO$(3)$
partial wave expansion in terms of the cross-channel angular momentum $J$. In particular,
in the high energy region Regge theory was extensively employed.
This implies that the high energy asymptotic behavior
$s^{\alpha(t)}$ of the amplitude is determined by the
(leading) Regge trajectory $\alpha(t)$,
which depends on the momentum transfer squared
$t=(p_2-p_1)^2$. It is assumed that the corresponding partial wave amplitudes
are analytic functions of $J$.
Leading Regge behavior then originates from  moving poles in the complex $J$-plane.
Besides such moving poles there also might exist so-called {\it fixed pole} singularities
(see {\it e.g.} Chapter~I of Ref.~\cite{Alfaro_red_book}) which

\noindent
\textbullet\;
do not move  with the change of $t$;

\noindent
\textbullet\;
can not be revealed by means of the analytic continuation in $J$.

A $J=J_0$ fixed pole singularity may arise from a cross channel exchange with a
non-Reggeized (elementary) particle of spin $J_0$ in the cross channel
(or from a contact interaction term).
It is then manifest as the Kronecker-$\delta$ singularity
in the complex $J$-plane. Its $t$-channel quantum numbers might be exemplified
{\it e.g.} by means of the Froissart-Gribov projection
\cite{ Gribov1961,Froissart1961}.

To our best knowledge,  a
$J=0$
fixed pole
in the context of Compton scattering off proton
first arose in 
Ref.~\cite{Creutz:1968ds} by
M.J.~Creutz, S.D.~Drell, and E.A.~Pashos as a constant,
denoted here as
$C_\infty$,
in the Regge-pole representation of the real forward Compton scattering amplitude
\begin{eqnarray}
f_1(\nu) = \sum_{\alpha\neq0 \atop \alpha \le 1} \frac{\beta_\alpha \nu^\alpha}{4\pi}\;
\frac{-1-e^{-i \pi \alpha}}{\sin(\pi \alpha)} + C_\infty,
\label{fixed-pole-Creutz}
\end{eqnarray}
where the energy variable is $\nu=\frac{s-u}{4M}$.
The representation
(\ref{fixed-pole-Creutz})
is supposed to be valid for the high energy region, while for low energy the
Compton amplitude $f_1(\nu)$ is known to satisfy
the Thomson limit
\begin{eqnarray}
f_1(0)=\lim_{\nu\to 0} f_1(\nu) = - \frac{e_p^2}{4\pi M},
\label{Th_limit}
\end{eqnarray}
where
$e_p$
is  the electric charge and
$M$
is the proton mass.
Therefore, in a loose sense, the value of
$C_\infty$
in (\ref{fixed-pole-Creutz})
characterizes how much from the Thomson limit
(\ref{Th_limit})
survives in the high energy regime.
In Ref.~\cite{Damashek:1969xj}, employing the subtracted dispersion relation of
Gell-Mann, Goldberger, and Thirring \cite{GellMann:1954db},
this was equivalently formulated in a more abstractly mathematical manner as the
{\it $J=0$ fixed pole  sum rule} expressed in terms of an analytically regularized inverse moment
\begin{eqnarray}
C_\infty = f_1(0) - \frac{2}{\pi}\int_{\nu_{\rm thr}}^{(\infty)}\frac{d\nu}{\nu}
{\rm Im} f_1(\nu),
\label{inverse_moment-RCS}
\end{eqnarray}
where the absorptive part is given by the total photoabsorbtion cross section
${\rm Im} f_1(\nu) =  \frac{\nu}{4\pi}\sigma_T(\nu)$.
First attempts
\cite{Damashek:1969xj, Dominguez:1970wu}
to extract the  value of
$J=0$
fixed pole contribution at
$t=0$
from experimental measurements employing finite--energy sum rules based on
(\ref{inverse_moment-RCS})
found its value roughly consistent with the Thomson limit.

The manifestation of the $J=0$ fixed pole contribution for virtual
Compton scattering,
{\it i.e.} of the constant contribution in the high-energy asymptotic limit,
was the subject of a broad discussion in the early seventies. S.~Brodsky, F. Close and J. Gunion
\cite{Brodsky:1971zh, Brodsky:1972vv} provided field theoretical arguments in favor of
such contribution originating from local two-photon interaction corresponding at the diagrammatic level to the
so-called ``seagull'' diagrams. A. Zee in Ref.~\cite{Zee:1972nq}
argued that $J=0$ fixed pole is
an inherent consequence of scaling behavior of the Compton amplitude in the Bjorken limit.
However, this reasoning was criticized by M.~Creutz \cite{Creutz:1973zf},
who disclaimed the existence of any theoretical argument in favor of such singularity independent of
specific models.

The importance of a $J=0$
fixed pole contribution has been anew emphasized more recently by  S.~Brodsky, F.~J.~Llanes-Estrada and A.~Szczepaniak
\cite{Brodsky:2008qu}.
They argue that  this contribution  possesses  unique features that are absent in amplitudes of other
processes such as meson production:

\noindent
\textbullet\;
The $J=0$ fixed pole contribution is a $t$-dependent constant
that is {\it independent} on the photon virtualities and is therefore universal.

\noindent
\textbullet\;
In the parton model its value is given by the inverse moment of the corresponding
$t$-dependent parton distribution function (PDF).

On the other hand, within the partonic picture, the subtraction constant,
which appears in the transverse non-flip DVCS amplitude, originates from the
so-called $D$-term.
Originally, the $D$-term was introduced in
Ref.~\cite{Polyakov:1999gs}
as a separate  addendum to a generalized parton distribution (GPD) that complements the polynomiality condition
for the unpolarized charge even GPD
$H^{(+)}$ within the double distribution representation
\cite{Mueller:1998fv,Radyushkin:1997ki}.
The existence of the $D$-term also has been justified from chiral dynamics.
The first Mellin moment of the $D$-term contributes into the
hadronic matrix elements of both the quark and gluon parts of the QCD energy momentum tensor.
The negative value of this specific moment has been argued to be a necessity for the
stability of the nucleon \cite{Polyakov:2002yz}.
It was realized that the $D$-term can be implemented as inherent part
of GPD within the modified double distribution representation
\cite{Belitsky:2000vk,Teryaev:2001qm,Radyushkin:2011dh}.
The $D$-term also turns to be a natural GPD ingredient within
the GPD representation based on the double partial wave expansion (in conformal and in
the cross-channel SO$(3)$ partial waves). This representation is known
in two versions (the approach based on the Mellin-Barnes integral techniques of Ref.~\cite{Mueller:2005ed},
and the so-called dual parametrization approach
\cite{Polyakov:2002wz,Polyakov:2008aa,SemenovTianShansky:2010zv})
that were recently found to be
completely equivalent
\cite{Muller:2014wxa}.
Within this approach it was first realized that the problem of universality of
$J=0$
fixed pole is related to the analytic properties of GPD moments in the complex conformal
spin $j$.
The analyticity assumption requiring the absence of
$j=-1$
fixed pole singularity in the Mellin space of
spectral functions allows to
express the subtraction constant through the analytically regularized inverse moment
sum rule and turns to be equivalent to the
$J=0$
fixed pole conjecture of
Ref.~\cite{Brodsky:2008qu}.

In this paper we restrict ourselves to Compton scattering in the generalized
Bjorken limit and provide 
a pedagogical presentation on the issue of
$J=0$
fixed pole conjecture and the $D$-term representation.
In Sec.~\ref{sec:DispRel} we review the
derivation of fixed-$t$ dispersion relations for the
Compton amplitude. We introduce a pair of equivalent
dispersion relations: the standard subtracted one and the
analytically regularized one. This
provides the $J=0$ fixed pole sum rule in terms of the analytically
regularized  inverse moment.
In Sec.~\ref{sec:deeply_virtual}
we employ these findings within the parton model to express  the corresponding sum rule in terms
of GPDs.  We discuss the mathematical subtleties 
in taking the high energy limit of  the
$D$-term sum rule. In Sec.~\ref{sec:DD} we show that the $J=0$ fixed pole conjecture holds true
if the $D$-term is an inherent part
of the GPD. This statement is illustrated with a toy
GPD model example in Appendix~\ref{AppA}.
Finally, in Sec.~\ref{sec:conclusions} we draw our conclusions.

\section{Dispersion approach for Compton scattering}
\label{sec:DispRel}
\setcounter{equation}{0}

\subsection{Subtracted and unsubtracted dispersion relations for Compton amplitude }

To parameterize the photon helicity amplitudes of Compton scattering
(\ref{ComptonScat})
we adopt the notations and conventions of
Ref.~\cite{Belitsky:2012ch}.
In particular, the transverse non-flip photon helicity amplitude reads:
\begin{eqnarray}
{\cal T}_{++} &\!\!\!=\!\!\!& \overline{u}(p_2)\Bigg[ \frac{\slash\!\!\!\! P}{P\cdot q}{\cal H}(\nu,t|Q_1^2,Q_2^2)  +i\sigma^{\alpha \beta}  \frac{P_\alpha \Delta_\beta}{2 M \, P\cdot q}
\\
&&\phantom{\overline{u}\Bigg[\frac{\slash\!\!\!\! P}{P\cdot q}} \times {\cal E}(\nu,t|Q_1^2,Q_2^2) \Bigg]u(p_1) + \mbox{parity odd part}\,,
\nonumber
\end{eqnarray}
where $\Delta = p_2-p_1$ and $\nu$ stands for the 
energy variable
\begin{equation}
\nu = \frac{P\cdot q}{2 M} = \frac{s-u}{4 M},\quad\mbox{with}\quad P=p_1+p_2\,,\quad  q=\frac{1}{2}(q_1+q_2)\,.
\label{var1}
\end{equation}

In what follows we mainly focus on the Compton form factor (CFF) ${\cal H}(\nu,t|Q_1^2,Q_2^2)$,
the analog of the Dirac form factor, which has even signature (even parity and  even charge  conjugation parity), {\it i.e.}, it is symmetric under the interchange of $\nu \to -\nu$.

\noindent
\textbullet\;
In the forward kinematics ($Q_1^2=Q_2^2=\cQ^2$, $t=0$)
its imaginary part corresponds to the deep inelastic scattering  structure function $F_1$
\begin{equation}
{\rm Im} \, {\cal H}(\nu,t=0|Q_1^2=Q_2^2=\cQ^2)= 2\pi F_1 (\xB,\cQ^2),
\label{cffH2FT}
\end{equation}
where $ \xB = \frac{\cQ^2}{2 M \nu}$.

\noindent
\textbullet\;
For real Compton scattering it can be expressed through the transverse photoabsorbtion cross section
\begin{equation}
{\rm Im} \,  {\cal H}(\nu,t=0|Q_1^2=Q_2^2=0)= 4\pi M \, {\rm Im} \,  f_1(\nu) = M  \nu \sigma_T(\nu).
\label{cffH2f1}
\end{equation}

\begin{figure}[th]
\centerline{\includegraphics[scale=0.4]{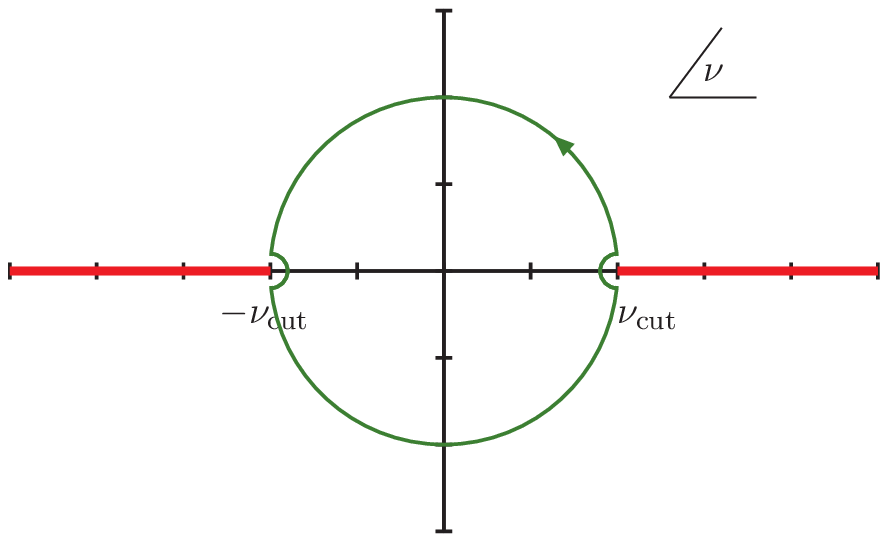}\hspace{1cm}
\includegraphics[scale=0.35]{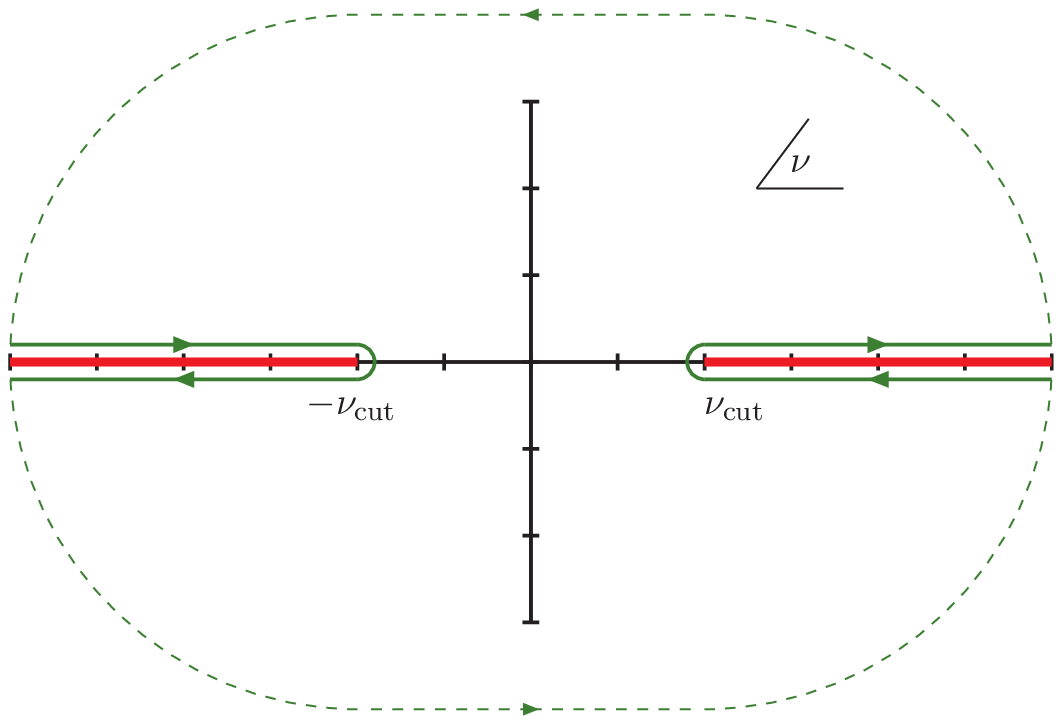}}
\caption{Left panel: Integration contour in the complex $\nu$ plane in eq.~(\ref{DR-0}).
 Right panel: Deformation of the integration contour in eq.~(\ref{UnsubtrDR}).
\label{fig:DR}}
\end{figure}
The derivation of fixed--$t$ dispersion relation (DR) for real photons or fixed space-like photon
virtualities is based on the
Cauchy theorem,
\begin{eqnarray}
{\cal H}(\nu,t|Q_1^2,Q_2^2)= \frac{1}{2\pi i} \oint d\nu^\prime \frac{1}{\nu^\prime -\nu} {\cal H}(\nu^\prime,t|Q_1^2,Q_2^2), \,
\label{DR-0}
\end{eqnarray}
(see left panel in Fig.~\ref{fig:DR}) and standard assumptions on
the analytic structure of the CFF.
In the following we concentrate  on the Bjorken limit. Therefore, the Born term can be safely neglected
and only the cuts along the real axis $[-\infty,-\nu_{\rm cut}]$ and $[\nu_{\rm cut},\infty]$,
which start at the pion production threshold,
$$\nu_{\rm cut}= \frac{Q_1^2+Q_2^2+t+(M+2m_\pi)^2-M^2}{4 M}$$
are to be accounted.
Deforming the integration  contour in
(\ref{DR-0})
as shown in the right panel of
Fig.~\ref{fig:DR}, and
assuming that
${\cal H}$
vanishes at infinity
($\lim_{|\nu|\to \infty} {\cal H}(\nu,t|Q_1^2,Q_2^2)= 0$), we work out the unsubtracted DR in the standard form,
\begin{equation}
{\cal H}(\nu,t|Q_1^2,Q_2^2)=
\frac{1}{\pi} \int^{\infty}_{\nu_{\rm cut}} d\nu^\prime \frac{2\nu^\prime\, {\rm Im} \, {\cal H}(\nu^\prime,t|Q_1^2,Q_2^2) }{\nu^{\prime 2} -\nu^2-i \epsilon}.
\label{UnsubtrDR}
\end{equation}
If
${\cal H}$
does not vanish at infinity,
the unsubtracted DR
(\ref{DR-0})
still formally provides the correct result once the contributions from the large semi-circles are retained.
However, it is practically of little use, since both the dispersive integral along the cuts and  the contribution from the large semi-circles are divergent.
Therefore, if one prefers to work with unsubtracted DR, {\it e.g.}, as done in
Ref.~\cite{Brodsky:2008qu},
it is indispensable to specify a regularization procedure at the point
$\nu=\infty$.

A possible choice, which was already briefly discussed in the Introduction, is the analytic regularization.
Here, the integration contour of the dispersive integral is deformed in a way that the integral along
the real axis is replaced by the loop integral in the complex plane that includes the point
$\nu=\infty$, denoted as $(\infty)$, for details see, {\it e.g.}, Ref.\ \cite{GelShi64}.
The unsubtracted DR
(\ref{DR-0})
then reads
\begin{eqnarray}
\label{DR-anareg-dv}
{\cal H}(\nu,t|Q_1^2,Q_2^2) &\!\!\!=\!\!\!& {\cal H}_\infty(t|Q_1^2,Q_2^2)
\\
&& + \frac{1}{\pi} \int^{(\infty)}_{\nu_{\rm cut}} d\nu^\prime \frac{2\nu^\prime\, {\rm Im} \, {\cal H}(\nu^\prime,t|Q_1^2,Q_2^2)}{\nu^{\prime 2}-\nu^2-i \epsilon},
\nonumber
\end{eqnarray}
where the constant
${\cal H}_\infty$,
arising from the analytic regularization at
$\nu=\infty$,
turns to be the analog of $C_\infty$ in the expansion of the real forward Compton scattering amplitude
(\ref{fixed-pole-Creutz}).
Within the Regge--pole expansion of the amplitude it is interpreted as the $J=0$ fixed pole contribution.

However, the analytically regularized DRs can be employed only once the analytic form of
the spectral function is explicitly known.
Therefore, the conventional form of  DR  employed within the deeply virtual  (d.v.) regime
is the subtracted DR with the subtraction taken at the unphysical  point
$\nu=0$:
\begin{eqnarray}
\label{DR-sub-dv}
{\cal H}(\nu,t|Q_1^2,Q_2^2) &\!\!\!\stackrel{\rm d.v.}{=}\!\!\!  & {\cal H}_0(t|Q_1^2,Q_2^2)
\\
&&+
 \frac{1}{\pi} \int^\infty_{\nu_{\rm cut}} \frac{d\nu^\prime}{\nu^\prime}
 \frac{2\nu^2\, {\rm Im} \, {\cal H}(\nu^\prime,t|Q_1^2,Q_2^2)}{\nu^{\prime 2} -\nu^2-i \epsilon} .
 \nonumber
\end{eqnarray}
The detailed derivation of (\ref{DR-sub-dv}) is given, {\it e.g.}, in Sec.~2.2 of
Ref.~\cite{Kumericki:2007sa}).

The dispersion relations (\ref{DR-anareg-dv}) and (\ref{DR-sub-dv})
are supposed to represent the same function. Therefore, the $J=0$
fixed pole contribution
${\cal H}_\infty$
could be related to the subtraction constant
${\cal H}_0$.
Plugging in the algebraic decomposition
$$
 \frac{2\nu^\prime}{\nu^{\prime 2} -\nu^2-i \epsilon}=   \frac{1}{\nu^\prime} \frac{2\nu^2}{\nu^{\prime 2} -\nu^2-i \epsilon}  + \frac{2}{\nu^\prime}
$$
of the Cauchy kernel  into (\ref{DR-anareg-dv}) and comparing it with (\ref{DR-sub-dv}), we read off the  sum rule
\begin{eqnarray}
\label{sumrule-J=0}
{\cal H}_\infty(t|Q_1^2,Q_2^2)&\!\!\!=\!\!\!&{\cal H}_0(t|Q_1^2,Q_2^2)
\\
&&-\frac{2}{\pi} \int^{(\infty)}_{\nu_{\rm cut}}  \frac{d\nu}{\nu}\, {\rm Im}{\cal H}(\nu,t|Q_1^2,Q_2^2)\,,
\nonumber
\end{eqnarray}
expressing the $J=0$ fixed pole contribution through the subtraction constant and the analytically regularized inverse moment of the absorptive part
of the amplitude.

\subsection{Dispersive approach in the scaling regime}

In general, the subtraction constant ${\cal H}_0(t|Q_1^2,Q_2^2)$ of the DR
(\ref{DR-sub-dv})
represents an unknown quantity. 
However, in the deeply virtual regime one can rely on the operator product expansion
and formulate the external principle allowing to fix the value of the
subtraction constant from the absorptive part.
In particular, within the leading twist-two approximation current conservation ensures
that for {\it equal} photon virtualities the subtraction constant vanishes:
${\cal H}_0(t|Q_1^2=Q_2^2=\cQ^2)\stackrel{\rm d.v.}{=}0$
(see Sec.~3.2.2 of Ref.~\cite{Kumericki:2007sa} for the detailed discussion),
while in the DVCS kinematics the subtraction constant corresponds to the $D$-term form factor
${\cal H}_0(t|Q_1^2=\cQ^2,Q_2^2=0)\stackrel{\rm d.v.}{=} 4{\cal D}(t)$.

Furthermore, in the framework of the operator product expansion it has been
conjectured in Ref.~\cite{Kumericki:2007sa} that in absence of the
$\delta_{j, -1}$ Kronecker singularity
(also called as $j=-1$ fixed pole contribution)
in the Mellin space of moments of the spectral function,
the subtraction constant
${\cal H}_0(t|Q_1^2,Q_2^2)$
for nonequal photon  virtualities
can be evaluated from the analytically regularized inverse moment
of the spectral function to leading twist accuracy to any order of perturbation
theory.
Within the convention used here, Eq.~(47) of Ref.~\cite{Kumericki:2007sa}
reads
\begin{eqnarray}
\label{sum-rule-D}
{\cal H}_0(t|Q_1^2,Q_2^2)  &\!\!\!=\!\!\!& \frac{2}{\pi} \int^{(\infty)}_{\nu_{\rm cut}} \frac{d\nu}{\nu}
  \\
&&\times{\rm Im}\left[
   {\cal H}(\nu,t|Q_1^2,Q_2^2) -{\cal H}(\nu,t|Q_1^2=Q_2^2)\right],
\nonumber
\end{eqnarray}
where the inverse $\nu$-moment is computed by the analytic continuation of $\nu$-Mellin moments.

Plugging this conjectured inverse moment sum rule
(\ref{sum-rule-D})
into the expression (\ref{sumrule-J=0})
for the $J=0$ fixed pole contribution,
one realizes that the
$J=0$
fixed pole is independent on the ratio of photon virtualities and can
be calculated from the equal photon virtuality case, yielding the conjecture of Ref.~\cite{Brodsky:2008qu}:
\begin{equation}
\label{J=0-conjecture}
{\cal H}_\infty(t|Q_1^2,Q_2^2)=
-\frac{2}{\pi} \int^{(\infty)}_{\nu_{\rm cut}} \frac{d\nu}{\nu}\, {\rm Im}{\cal H}(\nu,t|Q_1^2=Q_2^2)\,.
\end{equation}

Within the deeply virtual kinematics regime it is convenient 
to rewrite the DRs of the previous subsection in terms of scaling variables.
A natural choice is to use the Bjorken-like variable
$\xi$
and the skewness related scaling variable
$\eta$:
\begin{equation}
\xi = \frac{Q^2}{P\cdot q} = \frac{Q^2}{2 M \nu}; \,\qquad \eta = -\frac{\Delta\cdot q}{P\cdot q}=
-\frac{\Delta\cdot q}{2M\nu}\,,
\label{variable-scaling}
\end{equation}
where $Q^2=-q^2 \equiv -\frac{(q_1+q_2)^2}{4}$.
Here, instead of the scaling variable $\eta$,
we employ the photon asymmetry parameter
\be
\vartheta \equiv \eta/\xi = \frac{q_1^2-q_2^2}{q_1^2+q_2^2} + {\cal O}(t/Q^2),
\ee
which does not depend on the energy variable $\nu$ %
\footnote{
These variables are not
uniquely defined in the literature, {\it i.e.}, the various definitions differ by
$1/Q^2$
suppressed terms, that vanish in the generalized Bjorken limit.}.

\noindent
\textbullet\; For $t=0$,  the  $\vartheta=0$ case corresponds to the usual DIS
\phantom{\textbullet\;} kinematics.

\noindent
\textbullet\; The case $\vartheta=1$ corresponds to the  DVCS kinematics.

Within the scaling variables
(\ref{variable-scaling})
the analytically regularized DR
(\ref{DR-anareg-dv})
and  the subtracted one
(\ref{DR-sub-dv})
read as follows:
\begin{eqnarray}
{\cal H}(\xi,t|\vartheta) &\!\!\!=\!\!\!&  \frac{1}{\pi} \int^{1}_{(0)}
\frac{d\xi^\prime}{\xi^\prime} \frac{2\xi^2\, {\rm Im}{\cal H}(\xi^\prime,t|\vartheta)}{\xi^2-\xi^{\prime 2} -i \epsilon}
+  {\cal H}_\infty(t|\vartheta),\qquad
\label{DR-reg1}
\\
{\cal H}(\xi,t|\vartheta) &\!\!\!=\!\!\!& \frac{1}{\pi} \int^{1}_{0}
d\xi^\prime\frac{2\xi^\prime\, {\rm Im}{\cal H}(\xi^\prime,t|\vartheta)}{\xi^2-\xi^{\prime 2} -i \epsilon}
+  {\cal H}_0(t|\vartheta)\,.
\label{DR-sub1}
\end{eqnarray}
Here, the upper integration limit, given by
$\xi_{\rm cut} =\frac{Q^2}{2 M\nu_{\rm cut}}$,
has been set in the (generalized) Bjorken limit
to $\xi_{\rm cut}=1$
and the lower integration limit,
$\xi=0$,
corresponds to
$\nu=\infty$.
We emphasize that although the spectral function grows with increasing
$\xi^\prime$,
the analytically regularized DR
(\ref{DR-reg1})
can be evaluated so far its small-$\xi^\prime$
asymptotic is analytically known.
The equivalence of the two DRs
(\ref{DR-reg1}),
(\ref{DR-sub1})
is ensured by the sum rule
(\ref{sumrule-J=0}),
which  now reads
\begin{eqnarray}
{\cal H}_\infty(t|\vartheta) &\!\!\!=\!\!\!&   {\cal H}_0(t|\vartheta) -
\frac{2}{\pi} \int^{1}_{(0)} \frac{d\xi}{\xi}\,  {\rm Im}{\cal H}(\xi,t|\vartheta)\,.
\label{sum-rule-1}
\end{eqnarray}

\section{Dispersive versus pQCD approach}
\label{sec:deeply_virtual}
\setcounter{equation}{0}

In this section, within the GPD framework
set up in the familiar momentum fraction representation,
we point out the origin of the additional  fixed pole contribution
$\Delta{\cal H}_\infty$,
which eventually violates the $J=0$ fixed pole universality conjecture
(\ref{J=0-conjecture}).
In this approach, to the leading order (LO) accuracy, the CFF
${\cal H} (\xi,t|\vartheta)$
arises from the elementary amplitude
\begin{equation}
{\cal H} (\xi,t|\vartheta) \stackrel{\rm LO}{=}
  \int_{0}^1\!dx\,\frac{2x}{\xi^2-x^2- i \epsilon}
  H^{(+)} (x,\eta=\vartheta\xi,t)\,,
\label{eq:CFFLO}
\end{equation}
where
$H^{(+)}(x,\eta,t)= H(x,\eta,t) - H(-x,\eta,t)$
stands for the antisymmetric charge even quark GPD combination.
The imaginary part of the CFF is given by
the GPD value in the outer region
$\xi \ge \eta=\xi \vartheta $
for all allowed values
$|\vartheta| \le 1$,
\begin{eqnarray}
\frac{1}{\pi} {\rm Im}{\cal H} (\xi,t|\vartheta)
\stackrel{\rm LO}{=} H^{(+)}(x=\xi,\eta=\xi \vartheta,t)\,.
\label{ImH-LO}
\end{eqnarray}

Inserting the imaginary part (\ref{ImH-LO})
into the sum rule (\ref{sum-rule-1})
allows to express the $J=0$ fixed pole contribution
${\cal H}_\infty(t|\vartheta)$ to LO accuracy by
the GPD in the outer region:
\begin{equation}
{\cal H}_\infty(t|\vartheta) \stackrel{\rm LO}{=}
{\cal H}_0(t|\vartheta)
- 2 \int^{1}_{(0)} \frac{dx}{x}\, H^{(+)}(x,\vartheta x,t)\,.
\label{sum-rule-GPD}
\end{equation}
Now, by plugging the imaginary part (\ref{ImH-LO})
into the subtracted DR (\ref{DR-sub1})
and equating it with the LO convolution formula (\ref{eq:CFFLO}) for the CFF,
we obtain the GPD sum rule \cite{Kumericki:2008di},
which was originally worked out within the double distribution representation \cite{Teryaev:2005uj,Anikin:2007yh}:
\begin{equation}
4{\cal D}(t|\vartheta) \stackrel{\rm LO}{=}\int_{0}^1\!dx \frac{2x}{x^2-\xi^2}\!
 \left[\! H^{(+)} (x,\vartheta x,t) -H^{(+)} (x,\vartheta \xi,t)\right]
\label{D-sumrule}
\end{equation}
for the $D$-term form factor.
Note that for $\xi \ne 0$ the integrand in
(\ref{D-sumrule})
has an integrable singularity at
$x=\xi$.
The spectral function
$H^{(+)} (x,\vartheta x,t)$
has a branch point at
$x=0$,
while  the GPD
$H^{(+)} (x,\vartheta \xi,t)$
has a branch point at
$x=\vartheta \xi$
and vanishes at
$x=0$
due to antisymmetry in $x$.
The sum rule (\ref{D-sumrule}) is valid for all values of $\xi$
\footnote{For $|\xi|>1$, where one might rely on the GPD crossing properties \cite{Teryaev:2001qm}.},
where the special values $\xi\in \{0,1,1/\vartheta,\infty\}$ should be approached in special limiting procedures
(see Ref.~\cite{Kumericki:2008di} for a more detailed discussion).

\noindent
\textbullet\;
The low energy limit
$\xi\to \infty$
of
(\ref{D-sumrule})
is rather uncritical: the DR integral drops out and
$D$-term form factor is given in terms of the $D$-term
$d(x,t)$ that is here defined as a limiting value of the GPD:
\begin{eqnarray}
\label{D-LO}
4 {\cal D}(t|\vartheta) &\!\!\!\stackrel{\rm LO}{=}\!\!\!&
  \int_{0}^1\!dx\, \frac{2 
  x \vartheta^2}{1-x^2 \vartheta^2}\, d(x,t)
\\
\mbox{with} &&
 d(x,t)=\lim_{\xi\to\infty} H^{(+)} (\xi x,\xi,t)\,.
\nonumber
\end{eqnarray}

\noindent
\textbullet\; Contrarily, the high energy limit
$\xi \to 0$ of (\ref{D-sumrule}) requires special attention.
At a first glance this limit looks tempting to provide a proof
for the $J=0$ fixed pole conjecture of Ref.~\cite{Brodsky:2008qu}.
However, we would like to stress that
interchanging the integration and limiting procedure can render a wrong result,
since a squeezed contribution from the central GPD region might be missed.

Let us consider
the popular GPD representation in which the $D$-term
(denoted as $d^{\rm f.p.}$)
is an addenda that completes
polynomiality \cite{Polyakov:1999gs}:
\begin{equation}
H^{(+)}(x,\eta,t)=\HDD^{(+)}(x,\eta,t) + \theta(|\eta|-|x|) d^{\rm f.p.}(x/|\eta|,t)\,,
\label{Dt_added}
\end{equation}
where $\HDD^{(+)}$ has the rather common double distribution representation, see below Eq.\ \ref{DD2GPD} for $a=0$.
In the $N$-th Mellin moment
$\int_{-1}^1\! dx\,x^N \HDD^{(+)}(x,\eta,t)$ of the GPD
the highest possible power in
$\eta^{N+1}$ for odd $N$ is missing. We would like to show that the $D$-term addenda in
(\ref{Dt_added})
can be interpreted as the $J=0$ fixed pole contribution violating the
$J=0$ fixed pole sum rule conjectured in Ref.~\cite{Brodsky:2008qu}.
For simplicity, let us suppose that both the CFF spectral function
$\HDD^{(+)} (x,\vartheta x,t)$
and the GPD
$\HDD^{(+)} (x,\vartheta \xi,t)$
vanish  at
$x=0$,
allowing us to interchange safely the integration and
$\xi \to 0$
limiting procedure
in (\ref{D-sumrule}).
We plug the GPD (\ref{Dt_added}) into the $D$-term form factor sum rule
(\ref{D-sumrule}),
and separate  the integration region into the central,
$x \in [0,\vartheta \xi ]$,
and outer,
$x \in [\vartheta \xi,1]$,
GPD regions.
Then taking  the high energy limit
$\xi \to 0$,
we find that the corresponding sum rule reads
\begin{eqnarray}
\label{Dfp-exa}
4 {\cal D}(t |\vartheta ) &\!\!\! \stackrel{\rm LO}{=} \!\!\! & 2\int_{0}^1\!\frac{dx}{x}\!
 \left[\!
\HDD^{(+)}
(x,\vartheta x,t) -q^{(+)} (x,t)\right]\quad \\
&& + 4 {\cal D}^{\rm f.p.}(t| \vartheta),
\nonumber
\end{eqnarray}
where $q^{(+)}(x,t) \equiv \HDD^{(+)}(x,0,t)$ stands for the corresponding $t$-dependent PDF
(we assume that $q^{(+)} (x=0,t) =0$ to ensure convergence of the integral) and
\begin{eqnarray}
4{\cal D}^{\rm f.p.}(t|\vartheta)  &\!\!\!\stackrel{\rm LO}{=}\!\!\! & \lim_{\xi\to 0}
\int_{0}^{\vartheta \xi}\!dx\, \frac{2 x }{\xi^2-x^2 }\, d^{\rm f.p.}\!\left(\!\frac{x}{\vartheta \xi}\!,t\right)
\\
&\!\!\!\stackrel{\rm LO}{=}\!\!\! &  \int_{0}^{1}\!dx\, \frac{2 x \vartheta^2 }{1-\vartheta^2 x^2 }\, d^{\rm f.p.}(x,t)\,.
\nonumber
\end{eqnarray}
Note that since by construction the $d^{\rm f.p.}$ term
provides the complete contribution to the $D$-term form factor,
the inverse moment of the GPD/PDF combination in (\ref{Dfp-exa}) vanishes.

Now, inserting the $D$-term form factor (\ref{Dfp-exa}) into  (\ref{sum-rule-GPD}), we conclude that
in addition to the universal inverse  PDF  moment
the subtraction constant ${\cal H}_\infty(t|\vartheta)$
receives an additional non-universal contribution from the $D$-term
$d^{\rm f.p.}(x)$, defined solely within the GPD central region:
\begin{equation}
\label{H_infty-sum_rule}
{\cal H}_\infty(t|\vartheta) \stackrel{\rm LO}{=}
-2 \int^{1}_{0} \frac{dx}{x}\, q^{(+)}(x,t)+\underbrace{4 {\cal D}^{\rm f.p.}
(t|\vartheta)}_{\Delta{\cal H}_\infty(t|\vartheta)}\,.
\end{equation}
Note that the additional $J=0$ fixed pole contribution
$\Delta {\cal H}_\infty(t|\vartheta)$,
depends on the photon virtualities and is therefore non-universal.

\noindent
\textbullet\;
Therefore, we conclude that on general ground the GPD sum rule
(\ref{D-sumrule})
can not deliver a proof for the conjecture
(\ref{sum-rule-D}) that the subtraction
constant ($D$-term form factor) can be evaluated from an inverse moment of
the spectral function and so the
$J=0$
fixed pole
(\ref{J=0-conjecture})
universality conjecture remains also unproved.

\noindent
\textbullet\;
We also add that the high energy limit
$\xi \to 0$
and integration procedure in the GPD sum rule
(\ref{D-sumrule})
can not be interchanged in the presence of Regge behavior. Neglecting the central GPD region
now implies that one also throws away divergent terms that are needed to render a finite
$D$-term form factor result.

A particular example of a GPD model with a non-zero
$j=-1$
fixed pole contribution is provided by the calculation
\cite{SemenovTianShansky:2008mp}
of pion GPDs in the non-local chiral quark model
\cite{Praszalowicz:2003pr}.
In this model the universality conjecture
(\ref{J=0-conjecture})
is not valid due to a supplementary
$J=0$ fixed pole contribution
originating from the $D$-term $d^{\rm f.p.}(x,t)$,
which has to be added to make GPD satisfy the soft pion theorem
\cite{Pobylitsa:2001cz}
fixing pion GPDs in the limit
$\eta \to 1$.

Now we are about to spell out the relation between the $J=0$ fixed pole
contribution and the $D$-term form factor
${\cal D}(t|\vartheta)$
making special emphasize on the two kinds of analytical properties relevant for GPDs
and associated CFFs:

\noindent
\textbullet\; analyticity of CFFs in the cross channel angular momentum $J$;

\noindent
\textbullet\; analyticity of GPD Gegenbauer/Mellin moments in the variable $j$, labeling the conformal spin $j+2$
of twist-$2$ quark conformal basis operators %
\footnote{Here the covariant derivative $\stackrel{\leftrightarrow}{D}$
and the total derivative $\stackrel{\leftrightarrow}{\partial}$ are contracted with the light-cone vector $n$,
 $C_j^{\frac{3}{2}}$ stand for the usual Gegenbauer polynomials with the index $3/2$. }
\begin{equation}
{\cal O}_j^a
=
\frac{\Gamma(3/2)\Gamma(1+j)}{2^{j} \Gamma(3/2+j)}\, (i \!
\stackrel{\leftrightarrow}{\partial}_+)^j\; \bar{\psi} \lambda^a \gamma_+ \, C_j^{3/2}\!\left(\!
\frac{\stackrel{\leftrightarrow}{D}_{+}}{\stackrel{\leftrightarrow}{\partial}_+}\!
\right)\psi.
\label{Conf_basis}
\end{equation}

To deal with $J$ analytical properties of CFFs, following Sec.~6.3 of Ref.~\cite{Muller:2014wxa},
it is instructive to consider the Froissart-Gribov
projection \cite{Gribov1961,Froissart1961}
of the cross channel SO$(3)$ PWs of the CFF ${\cal H} (\xi,t|\vartheta)$:
\begin{equation}
a_{J}
(t|\vartheta) \equiv \frac{1}{2}\int_{-1}^{1}\!d(\cos\theta_t)\,
P_J(\cos\theta_t)  {\cal H}^{(+)}(\cos \theta_t,t|\vartheta),
\label{a_J(t)_def}
\end{equation}
where, neglecting the threshold corrections $\sim \sqrt{1- \frac{4M^2}{t}}$,
$$
\cos \theta_t= -\frac{1}{\vartheta \xi}+ {\cal O}(1/Q^2).
$$
For $J>0$ PWs the Froissart-Gribov projection provides
to LO accuracy
\begin{equation}
a_{J>0}(t|\vartheta) \stackrel{\rm LO}{=}
2 \int_0^1\! dx
\frac{{\cal Q}_J(1/x)}{x^2} H^{(+)}(x,\vartheta x,t)\,,
\label{Froissart-–Gribov}
\end{equation}
where ${\cal Q}_J(1/x)$ stand for the Legendre functions of the second kind.
For $J=0$ one obtains
\begin{eqnarray}
a_{J=0}(t|\vartheta) &\!\!\!\stackrel{\rm LO}{=}\!\!\!& 2 \int_0^1\! dx \left[\frac{{\cal Q}_0(1/x)}{x^2} -
\frac{1}{x}\right] H^{(+)}(x,\vartheta x,t)
\nonumber\\
&& + 4 {\cal D}(t|\vartheta)\,.
\label{Froissart-–Gribov_J=0}
\end{eqnarray}
Indeed, as clearly seen from Eqs.~(\ref{Froissart-–Gribov}) and
(\ref{Froissart-–Gribov_J=0}), the $J=0$ PW $a_{J=0}(t|\vartheta)$
might not be obtained from analytic continuation of
$a_{J>0}(t|\vartheta)$ to $J=0$. Therefore, analyticity in the cross channel angular momentum $J$
turns to be ``spoiled'' by the presence of a $J=0$ fixed pole contribution
\begin{equation}
a_{J=0}^{\rm f.p.}(t|\vartheta)
\stackrel{\rm LO}{=}  4 {\cal D}(t|\vartheta) -
2 \int^{1}_{(0)} \frac{d x}{x}\,  H^{(+)}(x,\vartheta x,t)\,.
\label{a_{J=0}^{f.p.}}
\end{equation}
Since the r.h.s.\ of Eqs.\ (\ref{sum-rule-GPD}) and (\ref{a_{J=0}^{f.p.}}) coincide,
one immediately recognizes that the constant ${\cal H}_\infty=a_{J=0}^{\rm f.p.}$ is
indeed the $J=0$ fixed pole contribution.

Note, that in the operator product expansion approach, {\it e.g.}, based on the conformal operator basis \cite{Kumericki:2007sa},
the presence of a $J=0$ fixed pole contribution (\ref{a_{J=0}^{f.p.}})  to the CFF ${\cal H}$ can be understood
form the absence of conformal operators with Lorentz spin $J\equiv j+1=0$.
Such a $j=-1$ contribution is effectively subtracted from the $J=0$ partial wave, see the $1/x$ moment in
the integral of Eq.\ (\ref{Froissart-–Gribov_J=0}). The analogous cancelation appears also in the
framework of dual parametrization of GPDs \cite{Muller:2014wxa}.

As pointed out in Refs.~\cite{Kumericki:2007sa,Muller:2014wxa},
the analytic properties in $j$ of GPD Gegenbauer/Mellin moments
control the validity of the internal duality principle for GPDs
(see also discussion in Ref.~\cite{Kumericki:2008di}). This principle relies on
the underlying Lorentz covariance and establishes the
relation between the inner and outer support regions for a GPD.
The absence of the $j=-1$ fixed pole contribution, violating analyticity in $j$, results
in a complete correspondence between the inner and outer GPD support regions.
This excludes the possibility to add a supplementary fixed pole $D$-term contribution
$d^{\rm f.p.}(x, \eta,t)$,
defined solely in the central GPD support region. In its turn, as explained above,
the absence of the $j=-1$ fixed pole $D$-term contribution leads to the
validity of  $J=0$ fixed pole universality conjecture of Ref.~\cite{Brodsky:2008qu}
(\ref{J=0-conjecture}):
\begin{equation}
{\cal H}_\infty(t|\vartheta)  \stackrel{\rm LO}{=}
-2 \int^{1}_{(0)} \frac{dx}{x}\, H^{(+)}(x,0,t).
\label{InvPdf}
\end{equation}
This statement is further illustrated within
the double distribution representation of GPDs in the next Section.

Moreover, we would like to emphasize that the inverse PDF moment (\ref{InvPdf})
can not be extracted from the $D$-term  form factor. The corresponding inverse moment is exactly
canceled within the GPD sum rule (\ref{D-sumrule})
for the $D$-term form factor.
This statement is obvious within the framework based on the conformal partial
wave expansion. Indeed, once the operator with the corresponding quantum numbers
($j=-1$, $J=0$)
does not appear within the conformal basis (\ref{Conf_basis}), the inverse PDF moment can not show up in the final
expression for the CFF.
Its somewhat artificial separation within the expression for the $D$-term form factor
(as the universal $J=0$ fixed pole contribution) suggests that it is exactly canceled
against the same term coming from the inverse moment of the absorptive part of the amplitude.
This issue is illustrated within the dual parametrization framework in
Sec.~6.2 of Ref.~\cite{Muller:2014wxa}. In other words, experimental data turn to be directly sensitive only
to a possible additional non-universal contribution
$\Delta {\cal H}_\infty(t|\vartheta)$
into ${\cal H}_\infty(t|\vartheta)$ ({\it c.f.} eq.~(\ref{H_infty-sum_rule})).

\section{$J=0$ fixed pole problem and GPD double distribution representation}
\label{sec:DD}
\setcounter{equation}{0}

According to the Mellin space analysis of
Ref.~\cite{Kumericki:2007sa}, a
$J=0$ fixed pole contribution originating from the
$D$-term should be absent if the $D$-term is the inherent part of a GPD.
To illustrate this statement, let us employ the double distribution (DD) representation
for the charge even GPD combination
$H^{(+)}(x,\eta)$
(for simplicity we omit the $t$-dependence and still adopt to a specific form of the DD representation)
\begin{eqnarray}
H^{(+)}(x,\eta) &\!\!\!=\!\!\!&
\int_{0}^{1}\!dy\int_{-1+y}^{1-y}\!dz\, \Big[(1-a x)\delta(x-y-z\eta)
\nonumber\\
&&\phantom{\int_{0}^{1}\!dy\int_{-1+y}^{1-y}\!dz} - \{x\to -x\} \Big] h(y,z)\,.
\qquad
\label{DD2GPD}
\end{eqnarray}
Here the  DD $h(y,z)$ is  symmetric in $z$
and antisymmetric in $y$. The factor
$(1-a x)$
is included in a way that for
$a=0$
the GPD polynomiality condition is not respected in its complete form (see
\cite{Polyakov:1999gs}),
while for
$a\neq0$
polynomiality is complete.
In the following we need to restrict
the admissible class of functions for the DD $h(y,z)$.
We assume that $h(y,z)$
has a `smooth' asymptotic behavior in the limit
$y\to 0$, in particular contributions concentrated in
$y=0$ ($\sim \delta(y)$
and its derivatives) are absent%
\footnote{To avoid confusions, let us add that the DD
representation is not uniquely defined and the
representation can be changed by a `gauge' transformation \cite{Teryaev:2001qm}.
In this way one can generate also representations with a
common $D$-term, however,  also the analytic properties
of the spectral function might be changed.  These
are more intricate technicalities since the GPD remains
unchanged under `gauge' transformations.}.
In order to employ the analytic regularization prescription
for the relevant integrals we need to specify
explicitly the analytic behavior of the DD for $y \sim 0$.
We assume the usual Regge-like behavior for DD
\begin{equation}
h(y,z) = \sum_{\alpha >0} y^{-\alpha}\, h_\alpha(z) + \big\{ {\rm terms \ \ regular \ \ at }\ \ {y \sim 0}\big\}
\end{equation}
with $h_\alpha(z) =  h_\alpha(-z)$.

The GPD spectral function
(\ref{ImH-LO}),
given by the GPD in the outer region, reads in terms of the DD as
\begin{equation}
H^{(+)}(x,\vartheta x) = (1-a x)
\int^{\frac{1-x}{1+\vartheta x}}_{\frac{1-x}{1-\vartheta x}}\!dz\, h([1+ \vartheta z] x ,z).
\label{h2H-spectral_function}
\end{equation}
For
$\vartheta=0$
it reduces to the corresponding ($t$-dependent) PDF,
\begin{equation}
q^{(+)}(x)= H^{(+)}(x,\vartheta x)\Big|_{\vartheta=0} = (1-a x)\int_{-1+x}^{1-x}\!dz\, h(x,z).
\end{equation}
The $D$-term form factor can be calculated from the limit
$\eta\to \infty$
(\ref{D-LO})
in which the
$y$-dependence in the
$\delta$-function drops out and only the
$a$
proportional term survives,
\begin{equation}
4{\cal D}(\vartheta)= -a\int_0^1\!dy\int_{0}^{1-y} \!dz\,
\frac{4 z^2 \vartheta^2}{1-z^2 \vartheta^2} h(y,z)\,.
\label{h2D}
\end{equation}

First, let us show that the $D$-term form factor sum rule (\ref{D-sumrule}) holds true for the DD representation (\ref{DD2GPD}). Plugging the latter  into the r.h.s.~of the former, we get
\begin{eqnarray}
&& \int_{0}^1\!dx\,
\frac{2x}{x^2-\xi^2}
\left[ H^{(+)} (x,\vartheta x,t) -H^{(+)} (x,\vartheta \xi,t)\right] 
\nonumber\\
&&  =\int_{0}^1\!dx \int_{0}^{1}\!dy\int_{-1+y}^{1-y}\!dz
\frac{2x(1-a x)}{x^2-\xi^2}
\nonumber\\
&&\phantom{\int_{0}^1\!dx \int}
\times \Big[\delta(x(1-\vartheta z)-y)-\delta(x-y-z \vartheta \xi)\Big]h(y,z)\,.
\nonumber
 \end{eqnarray}
Performing the $x$-integration and dropping in the resulting integrand
its antisymmetric part in $z$, which is proportional to $z h(y,z)$,
 \begin{eqnarray*}
&&
\frac{2 \vartheta  \, z h(y,z)}{(y+\xi )^2-(\vartheta  \xi z )^2}
\left[\xi-\frac{a  y (2\xi+ y)}{1-\vartheta ^2 z^2}\right]-\frac{2 a\vartheta ^2 z^2 }{1-\vartheta ^2 z^2}h(y,z)
\\
&&\Rightarrow  -\frac{2 a z^2 \vartheta ^2 }{1-\vartheta ^2 z^2 }h(y,z),
 \end{eqnarray*}
we immediately recover the $D$-term form factor expression
(\ref{h2D}) in terms of the DD.

Next, we calculate the inverse moment of the GPD spectral function in terms of the DD,
\begin{eqnarray}
\int_{(0)}^{1} \frac{dx}{x} H^{(+)}(x,\vartheta x) &\!\!\! = \!\!\! & \int_{(0)}^{1}\!dx\int_{0}^{1}\!dy\int_{-1+y}^{1-y}\!dz\,
\\
&&\times
\frac{1-a x}{x}\delta(x(1-\vartheta z)-y)h(y,z)\,,
\nonumber
\end{eqnarray}
where the small-$x$ behavior of the GPD spectral function inherits
the small-$y$ behavior of the DD. Therefore,
we regularize the
$y$
integral analytically, which allows to perform the $x$-integration.
This renders a well defined inverse moment in terms of the DD,
\begin{eqnarray}
\label{inverse-moment-DD}
\int_{(0)}^{1}\!dx\, \frac{2}{x} H^{(+)}(x,\vartheta x) &\!\!\! = \!\!\! & \int_{(0)}^{1}\!dy\int_{-1+y}^{1-y}\!dz\,
\\
&&\phantom{\int_{(0)}^{1}\!dy}\times\left[\frac{2}{y}  - \frac{2a}{1-\vartheta^2 z^2} \right]h(y,z)\,.
\nonumber
\end{eqnarray}
The  inverse moment of the DD in the r.h.s. of
(\ref{inverse-moment-DD})
can be rewritten employing the value of the inverse moment at $\vartheta=0$,
which yields
\begin{eqnarray}
\int_{(0)}^{1}\!dx\, \frac{2}{x} H^{(+)}(x,\vartheta x) &\!\!\! = \!\!\! & \int_{(0)}^{1}\!dx\, \frac{2}{x} q^{(+)}(x)
\\
&& -a \int_{0}^{1}\!dy\int_{0}^{1-y}\!dz\,
\frac{4\vartheta^2 z^2}{1-\vartheta^2 z^2} h(y,z)\,.
\nonumber
\end{eqnarray}
The second term on the r.h.s.\ is nothing but the $D$-term form factor
(\ref{h2D}) and, thus, we conclude that the sum rule
(\ref{sum-rule-D}) holds for the GPD (\ref{DD2GPD}).

Consequently, the $J=0$ fixed pole universality conjecture (\ref{J=0-conjecture})
[or equivalently (\ref{H_infty-sum_rule}) with ${\cal D}^{\rm f.p.} =0$] of
Ref.~\cite{Brodsky:2008qu}
is valid.
However, adding a separate $D$-term contribution
$d^{\rm f.p.}$
to the spectral representation
(\ref{DD2GPD})
\begin{equation}
H^{(+)}(x,\vartheta x) \to H^{(+)}(x,\vartheta x) + \theta(x\le \eta) d^{\rm f.p.}(x/\eta)
\label{DD+D_representation}
\end{equation}
leads to the breakdown of the
$J=0$
fixed pole universality conjecture, see equality (\ref{H_infty-sum_rule}), and results in the fixed pole
contribution into the $D$-term form factor which can not be computed from
the inverse moment of the GPD spectral function.

\section{Conclusions}
\label{sec:conclusions}

In this paper we addressed the $J=0$ fixed pole universality conjecture and
the related analyticity principle allowing to fix the
subtraction constant in the standard DR for the Compton scattering amplitude
from the absorptive part of the amplitude.
The latter, formulated within the
GPD framework by adopting the operator product expansion, holds true if a
$j=-1$
fixed pole singularity in Mellin space is absent.
This turns to be equivalent to the existence of the
GPD spectral representation in which the $D$-term is an inherent part of the GPD.
In this paper we reduced ourselves to considering the LO GPD framework,
although it was already demonstrated that the result is more general and is valid to all orders of perturbation theory.

In particular, we clarified that the
$J=0$
fixed pole universality conjecture can not be proven by merely taking the high energy limit of the
$D$-term sum rule (\ref{D-sumrule}). A $D$-term associated fixed pole contribution may arise
from a supplementary $D$-terms added in the central GPD region. This contribution is
overlooked by the naive version of the aforementioned limiting procedure. Generally, it may lead to breakup of
the $J=0$
fixed pole universality conjecture (\ref{J=0-conjecture}).

Instead, the relation between the $J=0$
fixed pole contribution and the $D$-term form factor only can be viewed as a manifestation of
equivalence between analytic properties of CFFs in the cross channel angular momentum $J$ and
the spectral properties of GPDs. Although the relevant analyticity principle ensuring the
validity of the $J=0$
fixed pole universality conjecture looks quite appealing, we can not provide reliable
theoretical arguments in its favor. Moreover, examples of field theoretical GPD models for which
this analyticity principle is violated are well known in the literature.

Therefore, we confirm our pessimistic conclusion from
\cite{Muller:2014wxa} that the absence of a $D$-term related $J=0$ fixed pole (or the
validity of the $J=0$ fixed pole universality conjecture) remains an external assumption,
which can probably never be proved theoretically.

In principle one may try to address the $J=0$ fixed pole universality conjecture
phenomenologically by verifying the GPD sum rule (\ref{D-sumrule})
for the $D$-term form factor.
This task certainly provides further motivation to build up a unique framework
for Compton scattering from real to the deeply virtual regime, launched in
\cite{Belitsky:2012ch}.
However, employing the GPD sum rule
for the $D$-term form factor requires the theoretical extrapolation of
experimental measurements into the high energy asymptotic regime.
This might imply a general problem, namely a phenomenological
test will be  biased by the theory framework and/or the utilized
model. Even the first step - the reliable extraction of the
$D$-term form factor from experimental data represents a considerable
challenge (see {\it e.g.} Ref.~\cite{Pasquini:2014vua}).

\section*{Acknowledgements}

K.S. is grateful to S.~Brodsky for the instructive and inspiring discussions during
LightCone 2013 meeting, to F.~J.~Llanes-Estrada for the useful correspondence, and to L. Szymanowski for the comments on the manuscript. The work was partly supported by
the French grant ANR PARTONS (ANR-12-MONU-0008-01) and  by the NRF of South Africa under CPRR grant no.~90509.

\setcounter{section}{0}
\setcounter{equation}{0}
\renewcommand{\thesection}{\Alph{section}}
\renewcommand{\theequation}{\thesection\arabic{equation}}

\section{A  toy GPD example }
\label{AppA}

To illustrate the general reasoning of Sec.~\ref{sec:DD}
we consider a simple toy GPD model that arises from the DD
\begin{equation}
h^{\rm toy}(y,z) = (N/2) y^{-\alpha},
\end{equation}
where
$N= M_2 \frac{\Gamma(5-\alpha)}{\Gamma(2-\alpha)[2+(1-a)(2-\alpha )]}$ is the convenient overall
normalization factor expressed in terms of the averaged parton momentum fraction
$M_2$, see below (\ref{toy-q}). We take $x$ and $\eta$
to be positive and restrict %
\footnote{The `pomeron' case $\alpha \sim 1$ can be treated in a similar fashion, however,
within the considered DD-representation  a residue function would be require
$h_\alpha(z)$
that vanishes at the boundary
$z=\pm 1$.}
ourselves to the case
$\alpha <1$.
For illustration we ambiguously add to the spectral
representation (\ref{DD2GPD})
a supplementary $D$-term  contribution
$d^{\rm f.p.}(x)$,
which vanishes at the boundaries
$ d^{\rm f.p.}(z=0) = d^{\rm f.p.}(z=1) =0 $.

The GPD is calculated from the DD-representation
(\ref{DD2GPD}), (\ref{DD+D_representation})
\begin{widetext}
\begin{eqnarray}
H^{\rm toy}(x,\eta|\alpha) &\!\!\!=\!\!\!& \frac{N}{2(1-\alpha ) \eta }
\Bigg\{\theta(x\le \eta) \left[(1-a x) \left(\frac{x+\eta }{1+\eta }\right)^{1-\alpha }
-(1+a x) \left(\frac{\eta -x}{1+\eta }\right)^{1-\alpha } \right]
\nonumber\\
&&\phantom{\frac{n}{2 (1-\alpha ) \eta }}+ \theta(x\ge\eta)(1-a x)
\left[\left(\frac{x+\eta }{1+\eta }\right)^{1-\alpha }-\left(\frac{x-\eta }{1-\eta }\right)^{1-\alpha }\right]
\Bigg\}
+ \theta(x\le \eta) d^{\rm f.p.}(x/\eta),
\label{H^{toy}}
\end{eqnarray}
\end{widetext}
For
$\eta>0$
the GPD vanishes at
$x=0$
and has branch points at
$x=\eta$
and
$x=1$.

\noindent
\textbullet\;
For $a\neq 0$ the polynomiality condition
is implemented in its full form irrespectively to the absence or presence of the fixed pole contribution.

\noindent
\textbullet\;  For $a=0$ the highest possible power of $\eta$ for a given Mellin moment of the GPD entirely arises from
$d^{\rm f.p.}$.

The GPD spectral function (\ref{ImH-LO}) is easily calculated from the GPD (\ref{H^{toy}})
by setting $\eta= x  \vartheta $ in the outer region
\begin{eqnarray}
\label{H-toy-spectral}
H^{\rm toy}(x,\vartheta x)&\!\!\!=\!\!\!&\frac{N x^{-\alpha } (1-a x) }{2\vartheta  (1-\alpha )}
\\
&&\times \left[\left(\frac{1+\vartheta }{1+x \vartheta }\right)^{1-\alpha }-\left(\frac{1-\vartheta }{1-x \vartheta }\right)^{1-\alpha }\right].
\nonumber
\end{eqnarray}
In particular, the PDF ($\vartheta =0$) and the GPD on the cross-over line  ($\vartheta =1$) read as following
\begin{eqnarray}
q^{\rm toy}(x)&=& N  x^{-\alpha } (1-a x)(1-x)\,,
\label{toy-q}
\\
H^{\rm toy}(x,x)&=&\frac{N}{1-\alpha}  \left(\frac{2x}{1+x}\right)^{-\alpha} \frac{1-a x}{1+x},
\end{eqnarray}
where  $\int_0^1\!dx\, x\, q^{\rm toy}(x) =M_2.$

The $D$-term consist of the integral GPD part, calculated from the low energy limit (\ref{D-LO}), and the fixed pole piece:
\begin{equation}
d^{\rm toy}(x) = -\frac{a N}{2(1-\alpha)} x (1-x
)^{1-\alpha} + d^{\rm f.p.}(x)\,.\qquad
\label{d-toy-spectral}
\end{equation}
Now, the $D$-term form factor (\ref{D-LO}) might be directly calculated by means of
the complete $D$-term (\ref{d-toy-spectral}), where
it contains the integral and the fixed pole part
\begin{eqnarray}
4{\cal D}^{\rm toy}(\vartheta) &\!\!\!=\!\!\!&  4{\cal D}^{\rm int}(\vartheta)
+ 4{\cal D}^{\rm f.p.}(\vartheta) \qquad \mbox{with}
\label{D-toy}
\\
4{\cal D}^{\rm int}(\vartheta)  &\!\!\!=\!\!\!& \frac{a N}{1-\alpha }\Bigg[\frac{2}{2-\alpha}+\frac{1-\vartheta}{(1-\alpha ) \vartheta }\;
{_2F_1}\left({1,1\atop 2-\alpha} \big|\vartheta\right)
\nonumber\\
&&\phantom{ \frac{a N}{1-\alpha } \Bigg[}-\frac{1+\vartheta}{(1-\alpha ) \vartheta }\;
{_2F_1}\left({1,1\atop 2-\alpha} \big|-\vartheta\right) \Bigg];
\nonumber
\\
 4{\cal D}^{\rm f.p.}(\vartheta)  &\!\!\!=\!\!\!&
\int_{0}^1\!dx\, \frac{4 x \vartheta^2}{1-x^2 \vartheta^2}\, d^{\rm f.p.}(x).
\nonumber
\end{eqnarray}
The individual contributions ${\cal D}^{\cdots}$ satisfy ${\cal D}^{\cdots}(\vartheta=0)=0 $.

The direct evaluation of the inverse moment from the GPD spectral function
(\ref{H-toy-spectral}) yields
\begin{equation}
\int_{(0)}^1\!dx\frac{2}{x} H^{\rm toy}(x,\vartheta x) =
-\frac{2 N}{1-\alpha}\left[\frac{1}{\alpha}+\frac{a}{2-\alpha}\right]+ 4{\cal D}^{\rm int}(\vartheta)\,.
\label{toy-inverse_moment}
\end{equation}
It contains a $\vartheta$ independent term and the $\vartheta$-dependence is entirely contained in the GPD
integral part of the $D$-term while  the fixed pole contribution is missing.

Consequently, the conjecture that the $D$-term form factor can be calculated from the inverse moment sum rule,
\begin{eqnarray*}
4{\cal D}^{\rm int}(\vartheta) &\!\!\! = \!\!\!& \int_{(0)}^1\!dx\frac{2}{x}\left[ H^{\rm toy}(x,\vartheta x)-H^{\rm toy}(x,0)\right]
\\
&\!\!\!\neq  \!\!\!& 4{\cal D}^{\rm toy}(\vartheta) =4{\cal D}^{\rm int}(\vartheta)
+ 4{\cal D}^{\rm f.p.}(\vartheta)
\end{eqnarray*}
is spoiled by the $D$-term related fixed pole contribution
${\cal D}^{\rm f.p.}(\vartheta)$.
In accordance with that, in the $J=0$ fixed pole
(\ref{sum-rule-GPD}), build from
the net $D$-term
(\ref{D-toy})
and the inverse moment
(\ref{toy-inverse_moment}),
only the GPD integral part of the $D$-term cancels out  while the fixed
pole related one induces a
$\vartheta$-dependence:
$$
{\cal H}_{\infty}(\vartheta) =4{\cal D}^{\rm f.p.}(\vartheta)+
\frac{2 N}{1-\alpha}\left[\frac{1}{\alpha}+\frac{a}{2-\alpha}\right]\,.
$$
Hence, our simple toy model with an ambiguous non-vanishing $D$-term related
fixed pole contribution  contradicts the conjecture of
Ref.~\cite{Brodsky:2008qu} that the
$J=0$
fixed pole is independent on the photon virtualities.


\end{document}